\documentclass[prl,twocolumn,showpacs,showkeys,superscriptaddress,preprintnumbers,floatfix]{revtex4}

\usepackage{url}
\usepackage{graphicx}
\usepackage{bm} 
\usepackage{amsfonts,amsmath,amssymb,amsbsy,latexsym}

\begin{document}

\markboth{Adaptive Dynamics for Interacting Markovian Processes}
{Sato, Ay}


\title{Adaptive Dynamics for Interacting Markovian Processes}

\author{Yuzuru Sato}
\email[Electronic address: ]{ysato@math.hokudai.ac.jp}
\affiliation{RIES, Hokkaido Univeristy, 
Kita 12 Nishi 6, Kita-ku, Sapporo 060-0812, Japan}

\author{Nihat Ay}
\email[Electronic address: ]{nay@mis.mpg.de}
\affiliation{Max Planck Institute for Mathematics in the Sciences, Inselstrasse 22, D-04103 Leipzig, Germany}
\affiliation{Santa Fe Institute, 1399 Hyde Park Road, Santa Fe, NM 87501, USA}

\begin{abstract}
Dynamics of information flow in adaptively interacting 
stochastic processes is studied. We give an extended form of game dynamics for 
Markovian processes and study its behavior to observe information flow 
 through the system. Examples of the adaptive dynamics for two 
stochastic processes interacting through matching pennies game
 interaction are exhibited along with underlying causal structure. 
\end{abstract}

\pacs{
  05.45.-a, 
  89.75.Fb  
  89.70.+c  
  02.50.Le, 
  }
\keywords{adaptive dynamics, information theory, game theory,
nonlinear dynamical systems} 


\maketitle



When studying the interaction and evolution of many stochastic processes that
are endowed with the ability to adapt to their enviroment, 
a natural question arises: how does information flow though the system and, 
moreover, how can we measure or calculate this information flow? 
From the viewpoint of large networks of stochastic elements, 
flow of information in the network has been studied \cite{Erb03, Ay03, Wen04, Ay06a}. 
In general, mutual information is not a representative 
measure of information flow in adaptive dynamics 
as its causal structure forms a complex network, making the 
 concept of information flow unclear. To address this problem, we 
 give an extended form of game dynamics for 
interacting Markovian processes and investigate information flow quantitatively. 

Suppose that $N$ stochastic processes $X_1,\ldots,X_N$ are interacting with 
each other. At each time step $\tau$, the unit $n$ sends 
a symbol $s_{n}\in\{0,1\}$ to the other units and receives at most $N-1$ symbols from the 
other units. We denote the global system state as $s=s_{1}\cdots s_{N}$. 
The next symbol sent by the unit $n$, $s_n'$, 
is dependent on the symbol received from the previous global state, $s$. 
Local transition probabilities for $n$-th unit are described as 
\begin{equation}
x^{(n)}_{{s_{n}}'|s}=P({X_{n}}(\tau+1)={s_{n}}'|X_{1}(\tau)=s_1, \ldots,
 X_N(\tau)=s_N), \nonumber
\end{equation}
where $n=1, \ldots, N$ and $x^{(n)}_{0|s}+x^{(n)}_{1|s}=1$. 
The transition probabilities 
$(x_{0|s}^{(n)},x_{1|s}^{(n)})$ is an element of a simplex denoted by
$\Delta^{(n)}_{s}$. 

We introduce a local adaptation process to change transition probabilities 
$x^{(n)}_{{s_{n}}'|s}$, assuming that adaptation is very slow compared 
with relaxation time of the global Markovian process. After the system
reaches a stationary state, each unit independently changes  
its stochastic structure by changing its transition probabilities. 
Assuming strong connectivity of the global Markovian kernel,
we study dynamics of transition probabilities in an ergodic subspace. 
This assumption corresponds to persistency of dynamics of 
transition probabilities $x^{(n)}_{s_{n}'|s}$ in the state space. 
Time evolution of $x^{(n)}_{s_{n}'|s}$ is driven by simple stochastic 
learning through interaction: 
reinforcements for transition probabilities of the unit $n$
to send $0$ and $1$ in the previous global state $s$ 
are given by the constants $a^{(n)}_{0|s}$ and $a^{(n)}_{1|s}$. 
The conditional expectation reinforcements $R^{(n)}_{{s_{n}}'|s}$ 
to chose each symbols ${s_{n}}'$ given the previous state $s$ 
are calculated with $a^{(n)}_{{s_{n}}'|s}$, $x_{{s_{n}}'|s}$, and the 
unique stationary distribution. For $X_{n}$, we give adaptive dynamics 
for probabilities of $s_{n}'$ given $s$ for $t\sim t+\Delta t$ 
\begin{equation}
x^{(n)}_{{s_{n}}'|s} (t+\Delta t)= \frac{x_{{s_{n}}'|s} 
 ^{(n)}(t)e^{\beta^{(n)} R_{{s_{n}}'|s} ^{(n)}(t)}} 
{\sum_{n}^{N} x_{{s_{n}}'|s}
 ^{(n)}(t)e^{\beta^{(n)} R_{{s_{n}}'|s} ^{(n)}(t)}}, 
\label{eqs:DiscreteLearningDynamics}
\end{equation}
where $\beta^{(n)}$ is the learning rate for the unit $n$.  
Here $\Delta t$ is much larger than the relaxation time of the global 
Markovian process. The continuous time model is given as 
\begin{equation} 
\frac{\dot{x^{(n)}_{{s_{n}}'|s}}(t)}{x^{(n)}_{{s_{n}}'|s}(t)}
=\beta^{(n)}(R^{(n)}(t)_{{s_{n}}'|s}-R^{n}_{|s}(t)),  
\label{eqs:LearningDynamics}
\end{equation}
for $n=1,\ldots, N$, where
$R^{(n)}_{|s}=\sum_{{s_{n}}'}x^{(n)}_{{s_{n}}'|s}R^{(n)}_{{s_{n}}'|s}$ 
is the conditional expectation of reinforcements over all possible 
symbols given the previous system state $s$. 
Intuitively, when $(R^{(n)}(t)_{{s_{n}}'|s}-R^{n}_{|s}(t))$ is positive, 
that is, the conditional expectation reinforcement for a symbol 
${s_{n}}'$ given $s$ is greater than the average of the expectation reinforcement given $s$, 
the logarithmic derivative of $x^{(n)}_{{s_{n}}'|s}(t)$ increases, and 
when negative, it decreases. The learning rate, $\beta^{(n)}$, controls 
the time scales of the adaptive dynamics of each unit $n$. 
(See \cite{Sat05} for the derivation of this model.) 
Note that Eq. (\ref{eqs:LearningDynamics}) represents adaptive 
dynamics with finite memories. 
Higher dimensional coupled ODEs are required for multiple Markovian
process and PDEs for non-Markovian process with infinitely 
long memories. 

\begin{figure}[htbp] 
\begin{center} 
\includegraphics[scale=0.4]{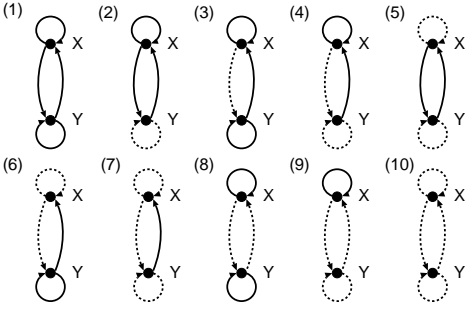} 
\end{center} 
\caption{Possible causal structure (case 1-10): $X\rightarrow Y$ indicates 
that Y receives symbols sent by X (information flow from X to Y). 
Dashed arrows indicate ignorance of received symbols (no information flow).} 
\label{fig:neta}
\end{figure}

Suppose that two biased coin tossing processes $X$ and $Y$ 
adaptively interact with each other. They produce a pair of symbols $ij$
at each time step, where $i$ and $j$ are either heads ($0$) or tails 
($1$). At the next time step, $X$ send a symbol $i'$ to $Y$ based on the 
previous pair of symbols $ij$, and vise versa. 
If there is a causal interaction with one step memory, the global stochastic 
process becomes a simple Markovian process. When $X$'s and $Y$'s behavior are 
causally separated, the whole system is a product of two biased
coin tossing processes (case 10 in Fig. \ref{fig:neta}). 

Considering Fig. \ref{fig:neta}, the extreme cases are 1 and 10. Case 1
corresponds to the situation, 
``each unit has one step memory of the previous global state $s$,'' and 
case 10 to, ``no information of $s$.'' Local transition probabilities of
$X$ and $Y$ are given as 
$(x_{i'|ij})=P(X'=i'|X=i,Y=j)$, and $(y_{j'|ij})=P(Y'=j'|X=i,Y=j)$, 
where $\sum_{i'}x_{i'|ij}=\sum_{j'}y_{j'|ij}=1$. The global Markovian
kernel is given with $(x_{i'|ij}y_{j'|ij})$ where $\sum_{i', j'} x_{i'|ij}y_{j'|ij}=1$. 
When $X$ and $Y$ match heads ($0$) or tails ($1$) of coins ($00$ or $11$), 
$Y$ reinforces the choice, and when they don't ($01$ or $10$), 
$X$ reinforces the choice. This interaction is called 
the matching pennies game in game theory. 
The reinforcements are given by a bi-matrix 
\begin{equation}
(A,B)=\left(  \left[\begin{array}{cc}   -\epsilon_X&\epsilon_X\\
    \epsilon_X &-\epsilon_X\\
  \end{array}\right],~~
\left[\begin{array}{cc}   \epsilon_Y&-\epsilon_Y\\
    -\epsilon_Y &\epsilon_Y\\
  \end{array}\right]
\right),
\label{eq:mp}
\end{equation}
where $0<\epsilon_X,\epsilon_Y<1$. The intereaction matrices, $A=(a_{ij})$ and $B=(b_{ji})$, are 
the reinforcements for $X$ and $Y$ for the global state $ij$. 
The Nash equilibrium of the game (\ref{eq:mp}) 
in terms of game theory is an uniformly random state $(1/2, 1/2)$. 
The conditional expectation reinforcements are given by 
$R^{X}_{i'|ij}=(A{\bf y}_{|ij})_{i'} ~{\rm and}~ R^{Y}_{j'|ij}=(B{\bf x}_{|ij})_{j'}$, 
where ${\bf x}_{|ij}=(x_{0|ij}, x_{1|ij})^T$, and ${\bf y}_{|ij}=(y_{0|ij}, y_{1|ij})^T$. 
Eq. (\ref{eqs:LearningDynamics}) reduces to 
\begin{eqnarray} 
\frac{\dot{x_{i'|ij}}}{x_{i'|ij}}
	& = & \beta^X[(A{\bf y}_{|ij})_{i'} - {\bf x}_{|ij}\cdot A{\bf y}_{|ij}],\nonumber\\
\frac{\dot{y_{j'|ij}}}{y_{j'|ij}}
	& = & \beta^Y[(B{\bf x}_{|ij})_{j'} - {\bf y}_{|ij}\cdot B{\bf	x}_{|ij}]. 
\label{eqs:LearningDynamics1}
\end{eqnarray}
Eq. (\ref{eqs:LearningDynamics1}) corresponds to adaptive dynamics for
an interacting Markovian processes in an 8-dimensional state space 
$\Pi_{i,j}\Delta_{ij}^{X}\times\Delta_{ij}^Y$, which is in the form of 
standard game dyamics. Similarly, for case 10, we have 
\begin{eqnarray} 
\frac{\dot{x_{i'|**}}}{x_{i'|**}}
	& = & \beta^X[(A{\bf y}_{|**})_{i'} - {\bf x}_{|**}\cdot A{\bf y}_{|**}],\nonumber\\
\frac{\dot{y_{j'|**}}}{y_{j'|**}}
	& = & \beta^Y[(B{\bf x}_{|**})_{j'} - {\bf y}_{|**}\cdot B{\bf x}_{|**}], 
\label{eqs:LearningDynamics10}
\end{eqnarray}
where ${\bf x}_{|**}=(x_{0|**}, x_{1|**})^T$, and 
${\bf y}_{|**}=(y_{0|**}, y_{1|**})^T$. 
Here, the * indicates ignorance of received symbols. 
Eq. (\ref{eqs:LearningDynamics10}) is, again,  
standard game dynamics in a 2-dimensional state space $\Delta^X\times\Delta^Y$. 
It is known that the dynamics of Eq. (\ref{eqs:LearningDynamics10}) is Hamiltonian
with a constant of motion $H=1/\beta^XD({\bf x}^*||{\bf x})+1/\beta^YD({\bf y}^*||{\bf y})$, where 
$D$ is Kullback divergence, and where $({\bf x}^*, {\bf y}^*)$ is the Nash equilibrium of the game $(A, B)$. 
The dynamics are neutrally stable periodic orbits for all 
range of parameters $\epsilon_X, \epsilon_Y$ \cite{Hof88, Hof96}. 
When the degree of freedom of the Hamiltonian systems is more than 2, 
and the bi-matrix $(A, B)$ gives asymmetric cyclical interaction, 
the dynamics can be chaotic \cite{Sat02, Sat03, Sat05}. Summarizing, 
if all units have complete information of the previous global state $s$ (case 1), 
or they are all causally separated with no information of $s$ (case 10),
we have a family of standard game dynamics given by
Eqs. (\ref{eqs:LearningDynamics1}) and (\ref{eqs:LearningDynamics10}).  

For intermediate cases $2 - 9$, showing in
Fig. \ref{fig:neta}, where units have partial information of $s$, 
we have explicit stationary distribution terms 
in the adaptive dynamics. Assuming the process is ergodic, $0<x_{i'|ij}, y_{j'|ij}<1$, an unique 
stationary distribution $(p(i,j))$ exists. We denote the marginal
stationary distributions ${\bf p}^X=(P(X=0), P(X=1))^T$, 
${\bf p}^Y=(P(Y=0), P(Y=1))^T$. 
The conditional stationary distribution of $i$, given the previous state
$j$, is denoted as 
$p(i|j)=p(i,j)/p(j)$, and those of $j$, given the previous state $i$, as $p(j|i)=p(i,j)/p(i)$. 

For case 2, with $R^{X}_{i'|ij}=(A{\bf y}_{|i*})_{i'}$ and 
$R^{Y}_{j'|i*}=\sum_{j} p(j|i) (B{\bf x}_{|ij})_{j'}$,
Eq. (\ref{eqs:LearningDynamics}) reduces to 
\begin{eqnarray} 
\frac{\dot{x_{i'|ij}}}{x_{i'|ij}}
	& = & \beta^X[(A{\bf y}_{|i*})_{i'} - {\bf x}_{|ij}\cdot A{\bf y}_{|i*}],\\
\frac{\dot{y_{j'|i*}}}{y_{j'|i*}}
	& = & \beta^Y[(\sum_{j} p(j|i)B{\bf x}_{|ij})_{j'} - {\bf y}_{|ij}\cdot
	(\sum_{j} p(j|i)B{\bf x}_{|ij})]. \nonumber
\label{eqs:LearningDynamics2}
\end{eqnarray}
Similarly, for case 5, with $R^{X}_{i'|*j}=(A{\bf p}^X)_{i'}$ and 
$R^{Y}_{j'|i*}=(B{\bf p}^Y)_{j'}$, we obtain 
\begin{eqnarray} 
\frac{\dot{x_{i'|*j}}}{x_{i'|*j}}
	& = & \beta^X[(A{\bf p}^Y)_{i'} - {\bf x}_{|*j}\cdot A{\bf p}^Y],\nonumber\\
\frac{\dot{y_{j'|i*}}}{y_{j'|i*}}
	& = & \beta^Y[(B{\bf p}^X)_{j'} - {\bf y}_{|i*}\cdot B{\bf p}^X].
\label{eqs:LearningDynamics5}
\end{eqnarray}

\begin{figure}[htbp]
\begin{center}
\includegraphics[scale=0.48]{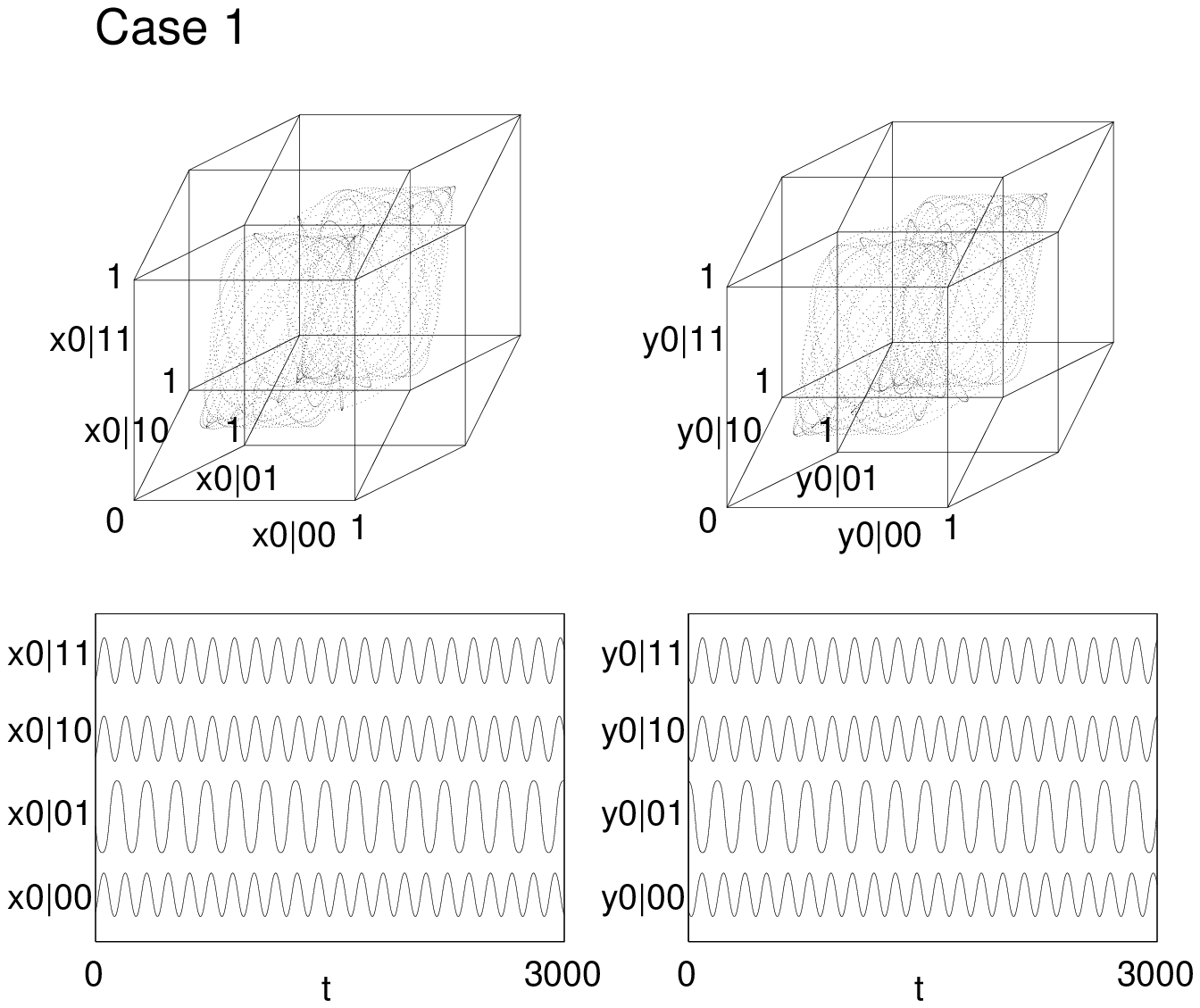}

\includegraphics[scale=0.48]{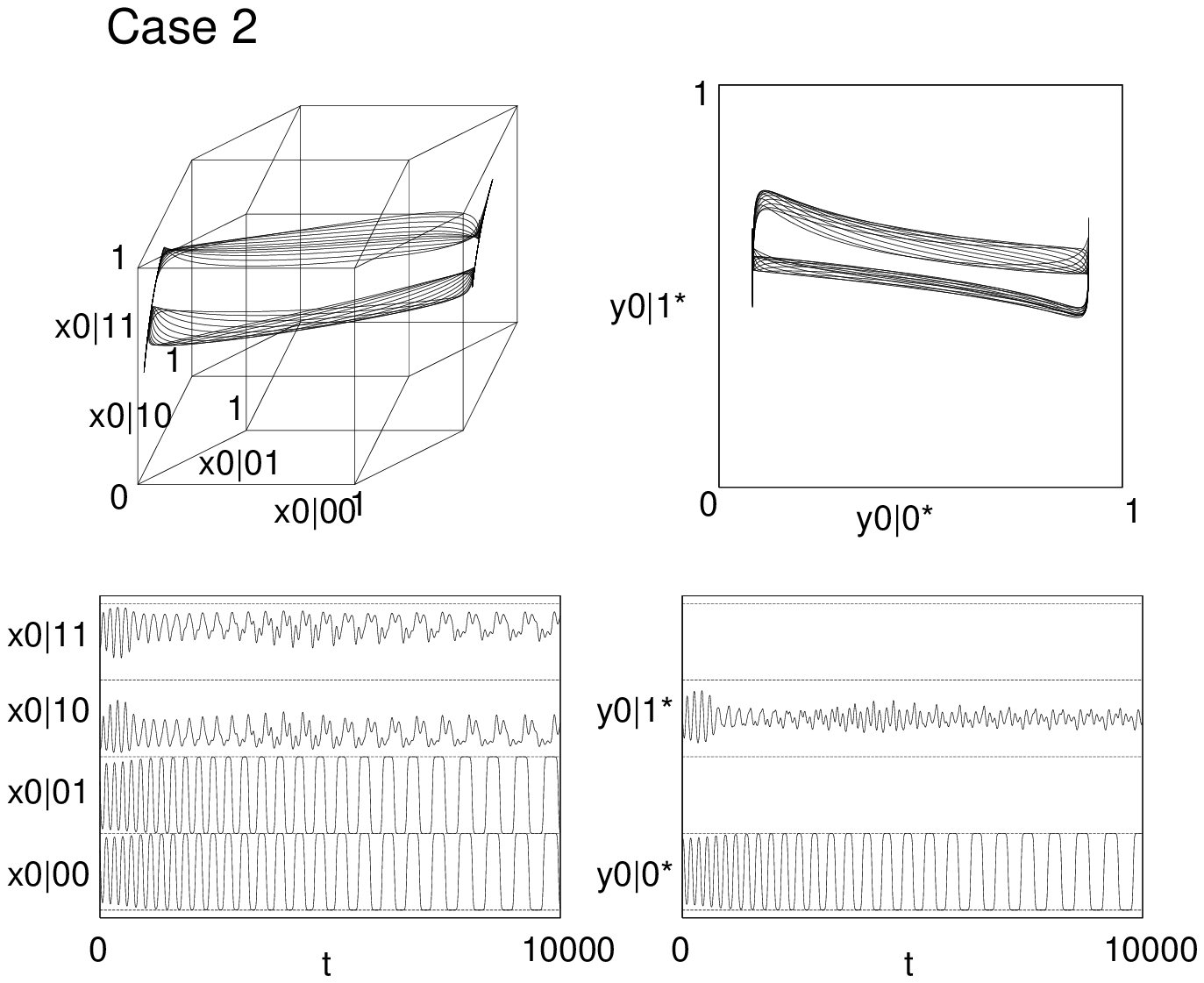}

\includegraphics[scale=0.48]{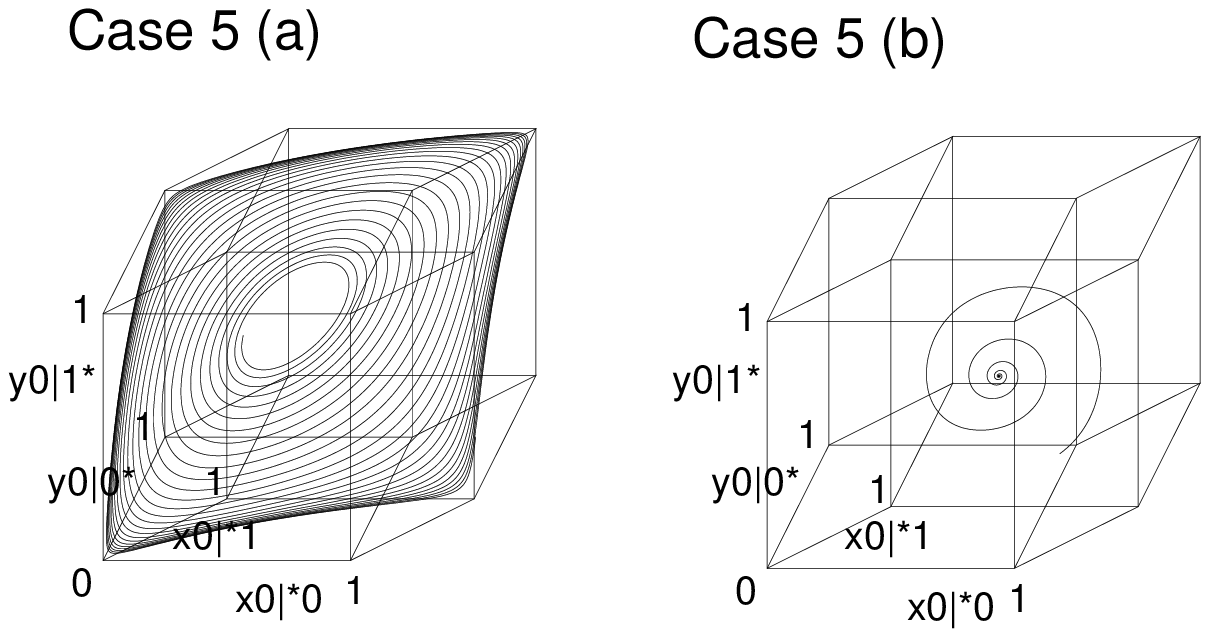}
\end{center}
\caption{(Top) Case 1:~ Neutrally stable quasi-periodic tori. 
(Middle) Case 2:~ A combination of quasi-periodic tori and 
transients to a heteroclinic cycle. 
(Bottom) Case 5:~ (a) Transients to a heteroclinic cycle which consists of vertex saddles 
$(x_{0|*0}, x_{0|*1}, y_{0|0*},  y_{0|1*})=$ $(0,0,0,0)$, $(1,1,0,0)$, $(0,0,1,1)$, $(1,1,1,1)$  
and (b) convergence to one of infinitely many neutrally stable fixed
 points which gives an uniform stationary distribution 
 $p(i,j)=\frac{1}{4}$; (in this case, converging to $(x_{0|*0}, x_{0|*1}, y_{0|0*}, y_{0|1*})=$ 
$(0.539057, 0.460943, 0.671772,  0.328228)$) are attracting sets.}
\label{fig:ld1}
\end{figure}

Note that $(p(i,j))$ are given as a function of $(x_{i'|ij})$ and 
$(y_{j'|ij})$, thus the equations of motion are in a closed form. 
For cases $2 - 9$, we have nonlinear couplings with a stationary
distribution, 
which is in contrast to the quasi-linear coupling of standard game dynamics. 
Eq.  (6) - (\ref{eqs:LearningDynamics5}) 
are both in an extended form of standard game dynamics. 

Let us now consider several examples. In examples, 
where the parameters are fixed to $\beta^X=\beta^Y$ and
$\epsilon_X=\epsilon_Y=0.5$, 
we have four types of dynamics: (1) neutrally stable periodic motion of Markovian kernel, 
(2) convergence to a fixed Markovian kernel that gives a uniform stationary
distribution, (3) sharp switching among almost 
deterministic Markovian kernel, (4) a combination of (1)-(3). 
In contrast to the matching pennies game dynamics which  
shows only neutrally stable periodic orbits, 
we obtain new types of dynamics naturally given by the Markovian structure. \\
{\bf Case 1 (Eq. (\ref{eqs:LearningDynamics1}))}:~ 
Neutrally stable quasi-periodic tori are observed.
They are simply a product of periodic orbits in the matching pennies game dynamics.
The dynamics of Eq. (\ref{eqs:LearningDynamics10}) is embedded in a 
subspace in the state space, given by 
$x_{i'|00}=x_{i'|01}=x_{i'|10}=x_{i'|11}$ and $y_{j'|00}=y_{j'|01}=y_{j'|10}=y_{j'|11}$.\\
{\bf Case 2 (Eq. (6))}:~ 
A combination of the dynamics of Eq. (\ref{eqs:LearningDynamics1}),
quasi-periodic tori, and the dynamics of
Eq. (\ref{eqs:LearningDynamics5}), transients to a heteroclinic cycle, are observed 
(Fig. \ref{fig:ld1}, middle). One of the 
infinitely many attracting periodic orbits corresponding to 
periodic orbits in Eqs. (\ref{eqs:LearningDynamics1}) 
is selected depending on initial conditions. \\
{\bf Case 5 (Eq. (\ref{eqs:LearningDynamics5}))}:~ 
Bi-stable dynamics is observed. A manifold which gives uniform stationary
distribution $p(i,j)=1/4$, is an attracting set. 
Fixed points on this attracting manifold are all neutrally stable. 
Heteroclinic cycles which consists of several vertex saddles are also 
attracting sets. Depending on initial conditions, 
either convergence to one of the fixed points on the attracting manifold 
or transients to one of the heteroclinic cycles are observed (Fig. \ref{fig:ld1}, bottom). 

In standard game dynamics which describes causally separated stochastic
process, information flow is always $0$. By using the Markovian extention of game dynamics, 
we can now quantify bi-directional information flow between stochastic units. 
Eq. (\ref{eq:si}) gives conditional mutual information of $Y$ and $X'$ 
given $Y$ and $X'$, which is a measure of stochastic dependence of $X'$
and $Y$ (sometime called transfer entropy, see \cite{Shaw84, Mat83, Kan86,
Sch00}). Recently, a new measure of information flow which describes deviation 
of two random variables from causal dependence, is formulated by Ay 
and Polani \cite{Ay06a}. Information flow from $Y$ to $X'$, given 
$X$ and $Y$, is defined by Eq. (\ref{eq:ci}) as a measure of causal dependence.
\begin{eqnarray}
&&I(Y:X'|X,Y)\nonumber\\
&=&\sum_{i',i,j} p(i',i ,j) \log \frac{p(i'|i,j)}{\sum_j p(j|i)p(i'|i,j)}\nonumber\\
&=& -\sum_{i',i} p(i) (\sum_j p(j|i)x_{i'|ij})\log (\sum_j p(j|i)x_{i'|ij})\nonumber\\
&&+\sum_{i,j}p(i,j)[\sum_{i'}x_{i'|ij}\log (x_{i'|ij})].\\
\label{eq:si}
&&I(Y\rightarrow\ X'|X,Y)\nonumber\\
&=&\sum_{i',i,j} p(i)p(j)p(i'|i ,j) \log \frac{p(i'|i,j)}{\sum_j p(j)p(i'|i,j)}\nonumber\\
&=&-\sum_{i',i} p(i) (\sum_j p(j)x_{i'|ij})\log (\sum_j p(j)x_{i'|ij})\nonumber\\
&&+\sum_{i,j}p(i)p(j)[\sum_{i'}x_{i'|ij}\log (x_{i'|ij})].
\label{eq:ci}
\end{eqnarray}

In the case that $Y$ is a fixed information source, 
$(y_{0|00},y_{0|01},y_{0|10},y_{0|11})=(1,0,1,0)$, the 
dynamics (\ref{eqs:LearningDynamics1}) with $\beta^Y=0$ 
monotonically converges to an optimal 
$(x_{0|00},x_{0|01},x_{0|10},x_{0|11})=(0,1,0,1)$. 
The system state $s$ is either $00$ or $11$ and $X$ is always rewarded. In this case, 
\begin{eqnarray}
I(X:Y'|X,Y)&=&0, ~~~I(X\rightarrow Y'|X,Y)=0, \\
I(Y:X'|X,Y)&=&0, ~~~I(Y\rightarrow X'|X,Y)=\log 2. \nonumber
\label{eqs:iyx}
\end{eqnarray}
There is information flow from $Y$ to $X$ because $X$ receives symbols 
sent by $Y$ and extracts information from $Y$'s behavior. 
Thus, $X$ is not stochastically dependent on $Y$ but, is causally 
dependent on $Y$. The above measure defined by (\ref{eq:ci}) clearly captures this property. 
Thus, intuitively, we can say that $I(Y\rightarrow X'|X,Y)$ is a more 
appropriate measure of the information flow. 

\begin{figure}[htbp]
\begin{center}
\includegraphics[scale=0.48]{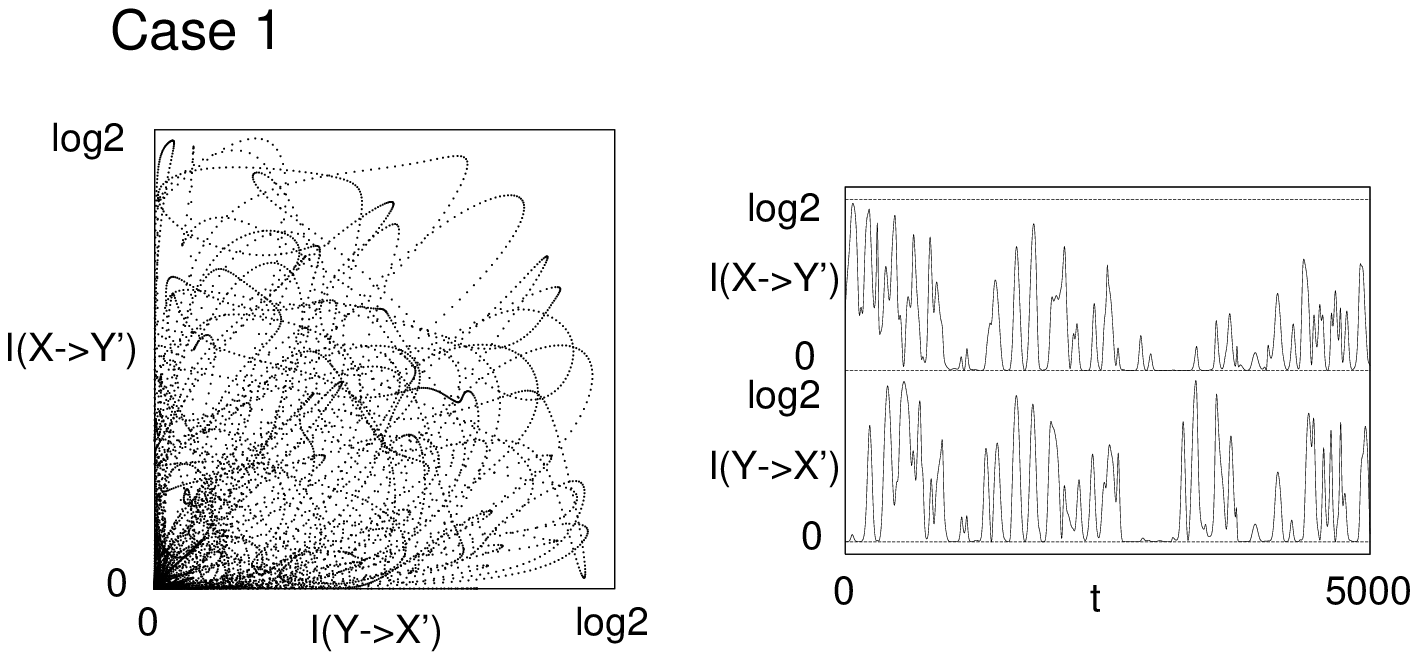}

\includegraphics[scale=0.48]{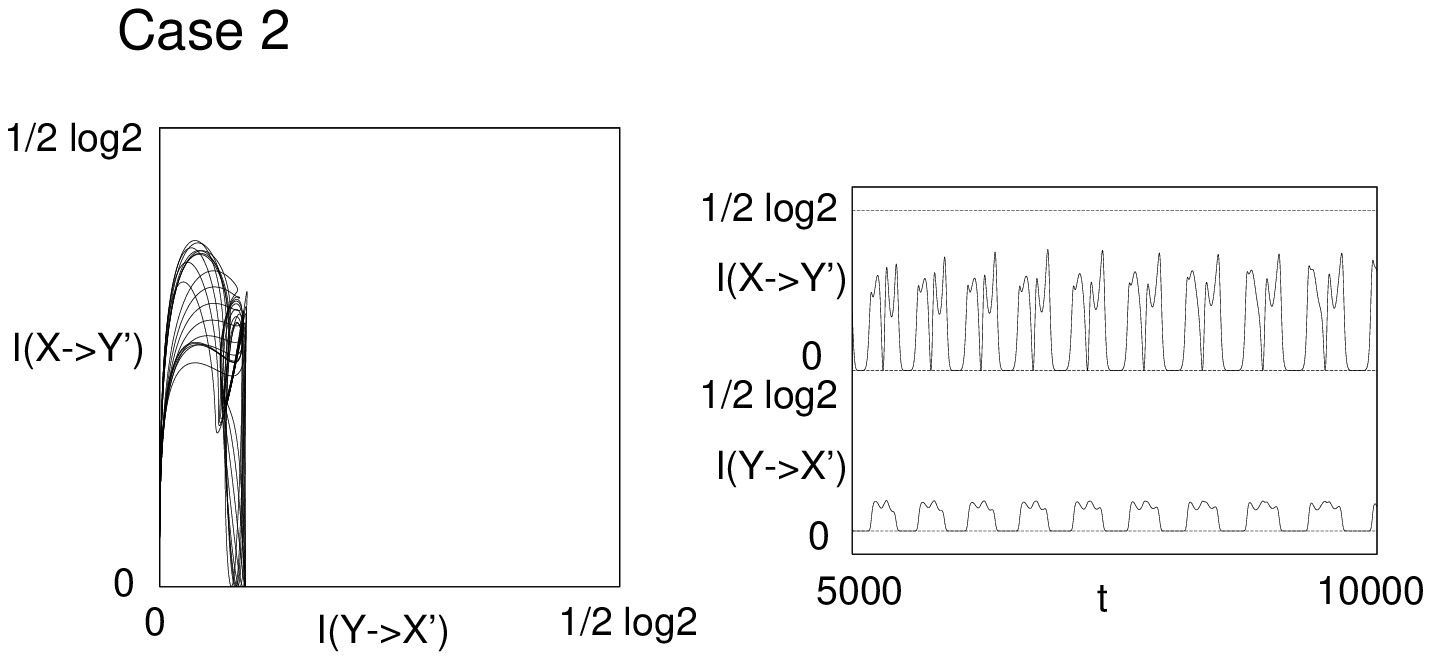}

\includegraphics[scale=0.48]{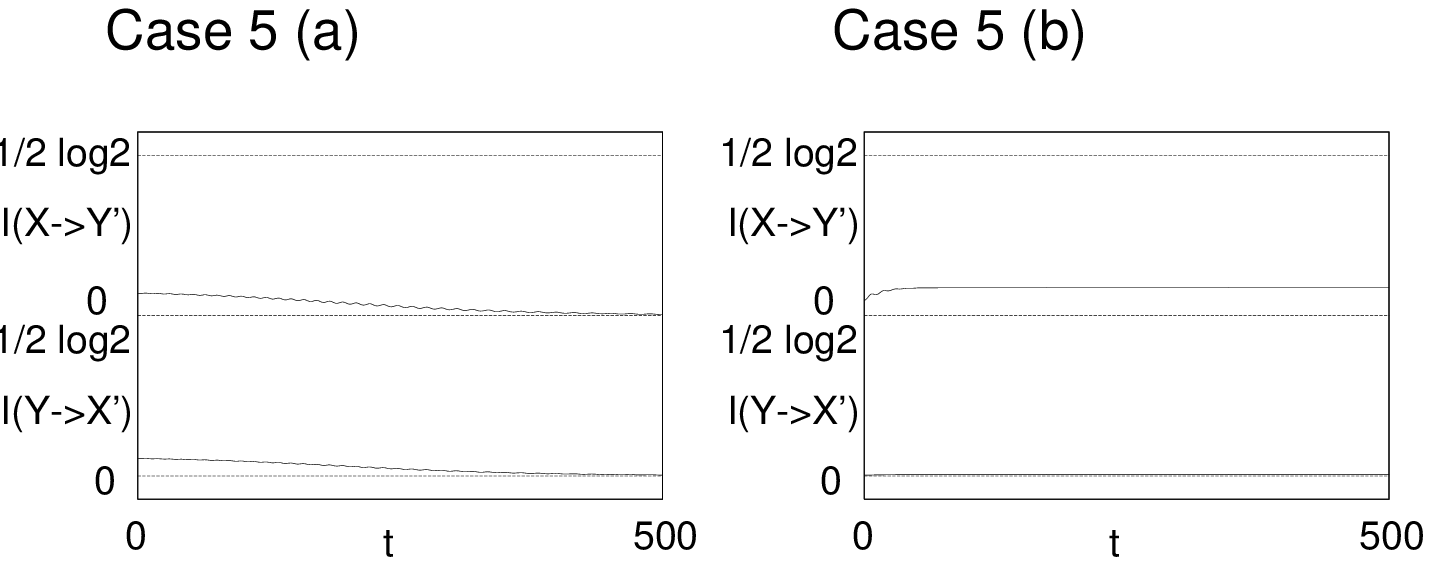}
\end{center}
\caption{(Top) Case 1:~ Aperiodic information flow. 
(Middle) Case 2:~ Periodic switching among aperiodic information flow. 
(Bottom) Case 5:~ (a) stationary information 
flow $I(Y\rightarrow X': X, Y)=I(X\rightarrow Y': X, Y)=0$. 
(b) stationary information flow 
$I(Y\rightarrow X': X, Y)=0.00305395$ and $I(Y\rightarrow X': X, Y)=0.06023025$.}
\label{fig:if}
\end{figure}

As shown in Fig. \ref{fig:if}, we observe (case 1)
aperiodic, (case 2) periodic switching among aperiodic, 
and (case 5) stationary information flow. 
In general, information flow vanishes 
when the system state is on a manifold $M_{0}$ defined by 
$x_{i'|ij}=\sum_{j} p(j) x_{i'|ij}$ and $y_{j'|ij}=\sum_{i}p(i) y_{j'|ij}$. 
Information flow is maximized to $\log 2$ when the system state is on a 
manifold $M_{1}$ defined by the set of points which have maximal
distance from $M_{0}$. Case 5 with bi-stability between a 
fixed point and heteloclinic cycle gives us a clear example of 
stationary information flow. Between the manifold $M_{0}$ and $M_{1}$ 
we have dynamic flow of information such as those in case 1 and 2 in 
Fig. \ref{fig:if}. Through adaptation, dynamic information flow emerges 
by keeping rewards as large as possible at each moment,   
and because of the complex game interaction and underlying causal structure. 

The above is an extention of game dynamics for interacting Markovian processes. 
If all units have complete information of the previous global state $s$,  
or they are all causally separated with no information of $s$, 
we have a family of standard game dynamics. 
For intermediate cases with partial information of $s$, 
we have explicit stationary distribution terms in 
the equations of motion. The presented examples 
show new types of phenomena in contrast to 
standard game dynamics. Dynamics of information flow between two units is discussed based on 
underlying causal structure. When units are ternary information sources, 
the presented game dynamics shows chaotic behavior even in 
the simplest case Eqs. (\ref{eqs:LearningDynamics10}) \cite{Sat02, Sat03, Sat05}. 
Studying adaptive dynamics for $N$ units with heterogeneous 
game interaction, and with various types of causal networks 
is left for a future work. 
Rigorous information theoretic analysis of the presented 
adaptive dynamics will be covered more elsewhere. 
The relationship between global and individual reward structure and
information flow among units 
would give us new insights in game theory. 
Applications to ecological and social dynamics, econophysics, 
and studies on learning in game are all straightforward. 

\vspace{1mm}
Authors thank D. Albers for careful reading of the manuscript, 
D. Krakauer and D. Polani for useful discussions. 
N. Ay thanks the Santa Fe Institute for support.  

\end{document}